\documentclass[]{aa}

\usepackage[dvips]{graphicx}
\usepackage{amsmath}
\usepackage{amssymb}

\begin{document}

\title{A multiwavelength study of the ultracompact HII region associated with IRAS
20178+4046}
\author{A. Tej\inst{1}, S. K. Ghosh\inst{1}, V. K.
Kulkarni\inst{2}, D. K. Ojha\inst{1}, R. P. Verma\inst{1} and S.
Vig\inst{1}\thanks{Present address: INAF-Osservatorio
Astrofisico di Arcetri, Largo E. Fermi, 51-50125, Italy}} 

\offprints{A. Tej, email: tej@tifr.res.in}

\institute{
Tata Institute of Fundamental Research, Homi Bhabha Road, Colaba, Mumbai 400 005, India\\
\and
National Centre for Radio Astrophysics, Post Bag 3, Ganeshkhind, Pune
411 007, India\\
}
\date{Received xxx / Accepted yyy}

\authorrunning{A. Tej et al.}

\titlerunning{Multiwavelength study of IRAS 20178+4046}

\abstract{}
{We present a multiwavelength study of the ultra compact HII region
associated with IRAS 20178+4046. This enables us to probe the different
components associated with this massive star forming region.}
{The radio emission from the ionized gas was mapped at 610 and 1280 MHz
using the Giant Metrewave Radio Telescope (GMRT), India.
We have used 2MASS $J H K_{s}$ data to study the nature of the
embedded sources associated with IRAS 20178+4046. 
Submillimetre emission from the cold dust at 450 and 850\,$\mu$m
was studied using JCMT-SCUBA.}
{The high-resolution radio continuum maps at 610 and 1280 MHz display
compact spherical morphology. The spectral type of the exciting source
is estimated to be $\sim$ B0.5 from the radio flux densities. However,
the near-infrared (NIR) data suggest the presence of several massive 
stars (spectral type earlier than O9) within
the compact ionized region.
Submillimetre emission shows the presence of two
dense cloud cores which are probably at different evolutionary stages.
The total mass of the cloud is estimated to 
be $\sim$ 700 -- 1500 $\rm M_{\odot}$
from the submillimetre emission at 450 and 850\,$\mu$m. 
}
{The multiwavelength study of this star forming complex reveals an interesting
scenario where we see the presence of different evolutionary stages in star formation.
The ultra compact HII region coinciding with the southern cloud core is at a
later stage of evolution compared to the northern core which is likely to be
a candidate protocluster.}

\maketitle

\section{Introduction}
\label{intro}
The early stages and evolution of massive stars are some of the 
least understood aspects of star formation. Ultracompact (UC) HII
regions are manifestations of newly formed massive (O or early B) stars deeply
embedded in the parental cloud.
IRAS 20178+4046 (G78.44+2.66) is a massive star forming region chosen
from the catalog of massive young stellar objects by Chan et al.
(1996). It has been classified as an UC HII region (Kurtz et al.
1994;1999). The kinematic distance estimates range from 1.5 kpc to 3.3
kpc (Kurtz et al. 1994; Caswell et al. 1975). In this paper we assume a
distance of 3.3 kpc. The far-infrared (FIR) luminosity from the IRAS fluxes
is estimated to be $\rm 7.0\times10^{4}\,L_{\odot}$ (Kurtz et al. 1994).
IRAS low-resolution spectra show a red
continuum beyond $\sim$ 13\,$\mu$m with the presence of the polycyclic
aromatic hydrocarbon (PAH) feature at 11.3\,$\mu$m (Volk \& Cohen
1989).
In their study of UC HII regions, Faison et al. (1998), present the
intermediate resolution infrared spectra of IRAS 20178+4046, where PAH
features at 3.3, 8.7 and 11.3\,$\mu$m, the [Ne II] line at 12.8\,$\mu$m
and the silicate feature at 9.7\,$\mu$m have been detected.
MSX mid-infrared (MIR) image at 21\,$\mu$m is presented by Crowther \&
Conti (2003). 

In a recent study, Verma et al. (2003) have mapped this
source in two FIR bands ($\sim$ 150 and 210\,$\mu$m) 
with the 1 m TIFR balloon-borne telescope. They also present the ISO
observations using the ISOCAM instrument in seven spectral bands
(3.30, 3.72, 6.00, 6.75, 7.75, 9.62 and 11.4\,$\mu$m). The
ISOCAM images mainly consist of a single component with a lobe 
towards south. They have modelled this source with a single core and
a dust density distribution of the form $r^{-2}$ comprising mainly of
silicates. 
IRAS 20178+4046 has been part of several maser
surveys (Codella et al. 1996; Baudry et al. 1997; Slysh et al. 1999;
Szymczak et al. 2000). No maser has been found to be associated with
this source. It has been detected in the CS(2--1) survey of
Bronfman et al. (1996). McCutcheon et al. (1991) and Wilking et al.
(1989) have observed this source in the CO lines as part of the survey
of protostellar candidates. 

In this paper, we present a multiwavelength study of this UC HII
region. In Sect.\,\,\ref{obvs.sect}, we describe the radio continuum
observations and the related data reduction procedure used. In
Sect.\,\,\ref{arch.sect}, we discuss other available datasets used in
the present study.
Section\,\,\ref{results.sect} presents a comprehensive
discussion of the results obtained. In Sect.\,\,\ref{starform.sect},
we discuss the general star forming scenario in the region associated
with IRAS 20178+4046 and in Sect.\,\,\ref{concl.sect} we
summarize the results.
\section{Observations and Data Reduction}
\label{obvs.sect}
\subsection{Radio continuum observations}
In order to probe the ionized gas component,
radio continuum interferometric mapping of the region around IRAS
20178+4046 was carried out using the Giant Metrewave Radio Telescope
(GMRT) array, India.  Observations were carried out at 1280 and 610 MHz.
The GMRT has a ``Y"-shaped hybrid configuration
of 30 antennas, each 45 m in diameter. There are six antennas along
each of the three arms (with arm length of $\sim$ 14 km). These
provide high angular resolution (longest baseline $\sim$ 25 km). The
rest of the twelve antennas are located in a
random and compact arrangement within $\rm 1\times1$ $\rm km^{2}$ near the
centre and is sensitive to large scale diffuse emission (shortest baseline
$\sim$ 100 m).
Details of the GMRT antennas and their configurations can be found in
Swarup et al. (1991). 
The radio sources 3C48 and 3C286 were used as the primary flux calibrators,
while 2052+365 was used as phase calibrator for both the 1280
and 610 MHz observations.

Data were reduced using AIPS. The data sets were
carefully checked using tasks UVPLT and VPLOT for bad data (owing to
dead antennas, bad baselines, interference, spikes, etc). Subsequent editing
was carried out using the tasks UVFLG and TVFLG. Maps of the field were
generated by Fourier inversion and subsequent cleaning using the task IMAGR.
Several iterations of self calibration were carried out to
obtain improved maps.
\section{Other available datasets}
\label{arch.sect}

\subsection{Near-infrared data from 2MASS}
Near-infrared (NIR) ($JHK_{s}$) data for point sources around IRAS
20178+4046 have been obtained from the Two Micron All Sky
Survey\footnote{This publication makes use of data products from the Two Micron All Sky
Survey, which is a joint project of the University of Massachusetts and the
Infrared Processing and Analysis Center/California Institute of Technology,
funded by the NASA and the NSF.}
(2MASS) Point Source Catalog (PSC). Source selection was based on the `read-flag' which gives the
uncertainties in the magnitudes. In our sample we retain only those
sources for which the `read-flag' values are 1 -- 3. The 2MASS data have been used to study
the sources associated with the UC HII region.

\subsection{Mid-infrared data from MSX}
The Midcourse Space Experiment\footnote{This research made use of data
products from the Midcourse
Space Experiment. Processing of the data was funded by the Ballistic
Missile Defense Organization with additional support from NASA Office of Space
Science. This research has also made use of the NASA/ IPAC Infrared Science
Archive, which is operated by the Jet Propulsion Laboratory, Caltech, under
contract with the NASA.}(MSX) surveyed the Galactic plane
in four MIR bands centered at 8.3 (A), 12.13
(C), 14.65 (D) and 21.34\,$\mu$m (E) at a spatial resolution of $\sim$
18$^{\prime\prime}$ (Price et al. 2001). The MIR data is used
to model the continuum SED from the interstellar medium around IRAS
20178+4046. Other than this, two of the MSX bands (A \& C) cover the
unidentified infrared bands (UIBs) at 6.2, 7.7, 8.7, 11.3 and 12.7\,$\mu$m. 
The MSX images in the four bands for the region around IRAS 20178+4046 have
been used to study the emission from the UIBs and to estimate the
spatial distribution of temperature and optical depth of the warm
interstellar dust. The MSX Point Source Catalog (MSX PSC) is used to study
MIR sources associated with the UC HII region.

\subsection{Sub-mm data from JCMT}
Submillimetre observations at 450 and 850\,$\mu$m using 
the Submillimetre Common-User Bolometer Array
(SCUBA) instrument of the James Clerk Maxwell Telescope\footnote{This paper
makes use of data from the James Clerk Maxwell Telescope Archive. The
JCMT is operated by the Joint Astronomy Centre on behalf of the UK particle
Physics and Astronomy Research Council, the National Research Council of
Canada and the Netherlands Organisation for Pure Research.} (JCMT) were retrieved
from the JCMT archives and processed using their standard pipeline SCUBA User
Reduction Facility (SURF). JCMT-SCUBA observations for the data used
in our study were carried out on 17 Aug 1998.
Uranus was used as the primary flux calibrator
for the maps. The submillimetre maps were used to study the cold dust 
environment associated with IRAS 20178+4046.

\section{Results and Discussion}
\label{results.sect}
\subsection{Radio continuum emission from ionized gas}
\begin{figure*}
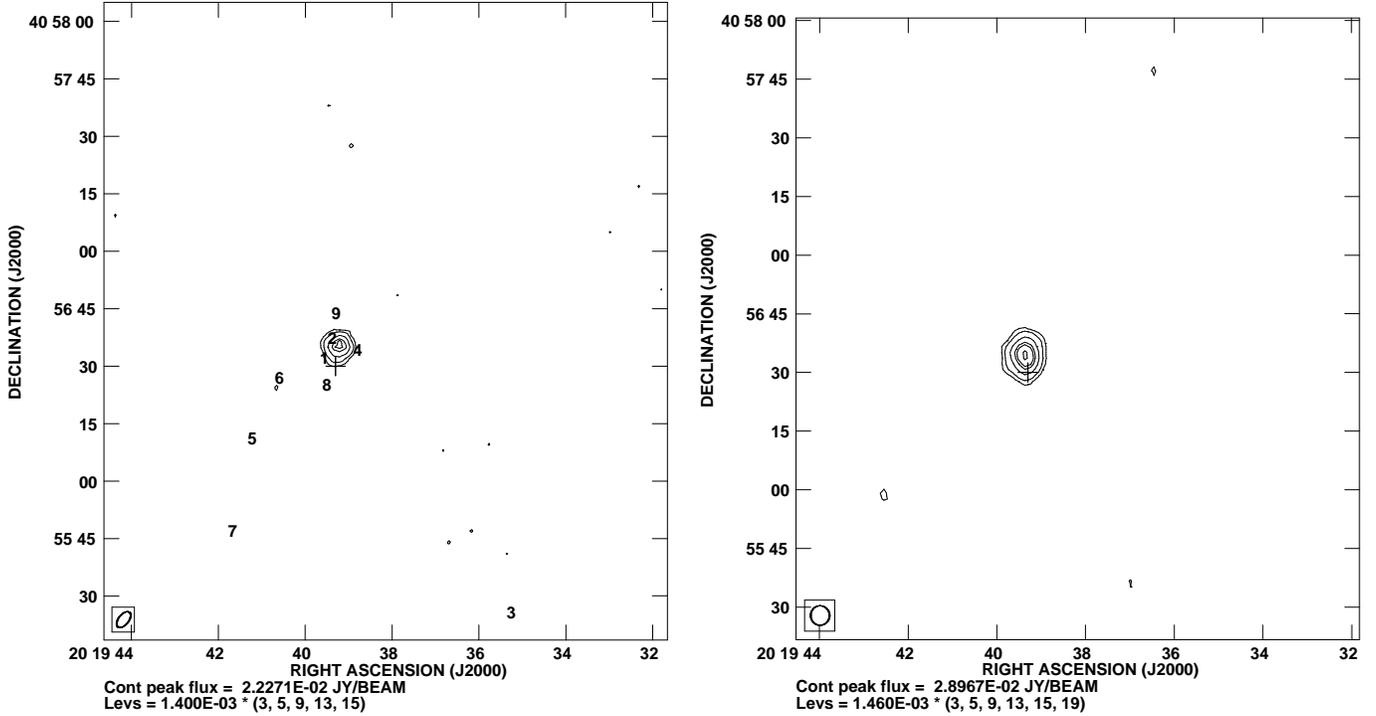

\centering
\resizebox{\hsize}{!}{\includegraphics{6462fig1l.ps}\includegraphics{6462fig1r.ps}}
\caption{High-resolution radio continuum maps 
at 1280 MHz (left) and 610 MHz (right) for the region around IRAS 20178+4046. 
The rms noises are $\sim$ 1.4 and 1.5 mJy/beam and the synthesized 
beam sizes are 4\arcsec.9 $\times$ 2\arcsec.7 and 5\arcsec.0$\times$4\arcsec.8 
at 1280 and 610 MHz, respectively. 2MASS sources with spectral type
estimations earlier than B0.5 (see Sect.\,\,\ref{nir}) are marked on the
1280 MHz map. The plus sign marks the position of the IRAS point
source.} 
\label{1280_610.fig}
\end{figure*}

High-resolution radio maps were generated from the GMRT data. 
The radio continuum emission from the ionized gas associated with the
region around IRAS 20178+4046 at 1280 and 610 MHz is shown in
Fig.\,\,\ref{1280_610.fig}. The details
of the observations and maps are given in Table\,\,\ref{radio.tab}.
For the 610 MHz observations, we present the map generated by using only
baselines larger than  1 kilo-lambda. 
Correction factors for the system temperature ($\rm T_{sys}$) were obtained from
additional observations of the neighbouring sky on and off the Galactic plane.
The correction factors are 1.2 and 1.6 for 1280 and 610 MHz, respectively. 
The generated maps were scaled using the 
$\rm T_{sys}$ correction factors obtained from our observations.
The integrated flux densities from our maps 
at 1280 and 610 MHz are 57$\pm$4 and 66$\pm$4 mJy, respectively. 
The flux densities are estimated by integrating down to the
lowest selected contour at the
3$\sigma$ level, where $\sigma$-s are the rms noises in the 
respective maps.
The corresponding size of the UC HII region is $\sim$ 10\arcsec and
15\arcsec at 1280 and 610 MHz, respectively.
Considering the errors, these flux densities 
are consistent with the hypothesis that the emission at these
frequencies are optically thin.
For this UC HII region, Wilking et al. (1989) estimate flux
density values of 80 and 65 mJy from their 15 and 5 GHz radio maps and
McCutcheon et al. (1991) obtain a value of 69 mJy at 5 GHz.
As expected from our low frequency GMRT data, the 5 GHz
flux density values of the above studies also indicate an 
optically thin continuum flux distribution.
The higher value at 15 GHz from Wilking et al. (1989)
may be due to the contribution from the extended emission. 

Kurtz et al. (1994;1999) have also studied this source at 
8.3 and 15 GHz.
Kurtz et al. (1994) estimate flux density values of
 82.2 and 35.6 mJy from their very high-resolution
($<$ 1\arcsec) maps using the VLA B array at 8.3 and 15 GHz, 
respectively. The integration
boxes used by them are $\rm 9\arcsec.4 \times 8\arcsec.9$ and 
$\rm 9\arcsec.5 \times 8\arcsec.5$ for the two bands, respectively. 
Further observations by Kurtz et
al. (1999) at 8.3 GHz with the VLA D array give a value 
of 68 and 158 mJy for 50\arcsec\, and
300\arcsec\, boxes, respectively with a synthesized beam size of $\rm 8\arcsec.7
\times 7\arcsec.8$. It should be noted here that the
300\arcsec\, box covers a much larger area and includes the
surrounding extended diffuse emission. 

\begin{table}
\caption{Details of the radio interferometric continuum observations
of the ionized region associated with IRAS 20178+4046.}
\label{radio.tab}
\begin{tabular}{l|c|c}
\hline\hline
  Details         & 1280 MHz & 610 MHz \\
\hline
Date of Obs.      & 2 Aug 2003 & 18 Sep 2004 \\
Primary beam      & 26\arcmin & 54\arcmin \\
Cont. bandwidth (MHz) & 16        & 16  \\
Synth. beam   &  4\arcsec.9 $\times$ 2\arcsec.7 &
5\arcsec.0$\times$4\arcsec.8$^\dagger$ \\
Position angle. (deg)  &  -38.6            & -0.25 \\
Peak Flux (mJy/beam)& 22.3 & 29.0  \\
rms noise ($\sigma$)(mJy/beam) & 1.4            & 1.5 \\
Largest detectable   & $\sim$ 8\arcmin & $\sim$ 17\arcmin\\
scale size           &                  &                 \\
Highest angular      & $\sim$ 2\arcsec  & $\sim$ 4\arcsec \\
resolution           &                  &           \\
Int. Flux density$^\ddagger$ (mJy) & 57$\pm$4       & 66$\pm$4 \\
\hline
\end{tabular}\\
\footnotesize $^\dagger$ For 610 MHz, the synthesized 
beam size is for the baselines $>$ 1 kilo-lambda\\
$^\ddagger$ Flux densities are obtained by integrating down to
the contour at 3$\sigma$ level, where $\sigma$ represents
the map noise in corresponding frequency band.\\
\end{table}

The contour maps at 1280 and 610 MHz generated from the GMRT observations 
display a simple compact spherical morphology.
In comparison, the high-resolution ($<$ 1\arcsec) radio map at 
8.3 GHz of Kurtz et al.  (1994) show a cometary UC HII region.
However, in their 15 GHz map, the sources are resolved out and
morphology looks `clumpy'. This is possibly the reason for
the low integrated flux density (35.6 mJy) obtained at this
frequency.
The lower resolution (8\arcsec.7 $\times$
7\arcsec.8) 8.3 GHz map by Kurtz et al. (1999) is unresolved with an
extended east-west emission seen to the south which according to
the authors is unlikely to be connected with the UC HII region. 
A point worth mentioning here is
that we possibly detect extended diffuse emission in the
low resolution map at 610 MHz (not presented in this paper).
But the signal-to-noise ratio of the map is not sufficient
to conclusively comment on the nature and association if 
any of this extended diffuse emission with the UC HII region.
However, we do not detect any such emission in our 1280 MHz
map. This could be because the flux levels of the extended
emission is below our sensitivity limit at 1280 MHz.

Using the low-frequency flux densities at 1280 and 610 MHz from our GMRT
observations and the 8.3 GHz data from Kurtz et al. (1999), we derived
the physical properties of the compact core of the HII
region associated with IRAS 20178+4046. Mezger \& Henderson (1967)
have shown that for a homogeneous and spherically
symmetric core, the flux density can be written as
\begin{equation}
S = 3.07 \times 10^{-2} T_{e} \nu^{2} \Omega (1 - e^{-\tau(\nu)})
\end{equation}
where
\begin{equation}
\tau(\nu) = 1.643a \times 10^{5} \nu^{-2.1} (EM) T_{e}^{-1.35}
\end{equation}
where $S$ is the integrated flux density
in Jy, $T_{e}$ the electron temperature in K, $\nu$ the
frequency of observation in MHz, $\tau$ the optical depth, $\Omega$
the solid angle subtended by the source in steradians, and $EM$
the emission measure in $\rm cm^{-6} pc$. Also, $a$ is a correction factor
and we used a value of 0.99 (using Table 6 of Mezger \& Henderson 1967)
for the frequency range 0.6 -- 8 GHz and $T_{e} = \rm 8000 K$.
The two GMRT maps were convolved to a common angular resolution of
$\rm 8\arcsec.7 \times 7\arcsec.8$, which is the resolution
of the 8.3 GHz map of Kurtz et al. (1999).
In our case, since the core is unresolved, $\Omega$ is taken as
this synthesized convolved beam size (i.e $\Omega = 1.133 \times \theta_{a} \times
\theta_{b}$, where $\theta_{a}$ and $\theta_{b}$ are the half power
beam sizes). The peak flux densities of the core in
the 0.6 -- 8.3 GHz frequency range are consistent with an optically
thin HII region. Assuming a typical electron temperature of 8000 K, these flux densities
were used to fit the above equations. The best fit value for the emission
measure is $\rm 2.9 \pm 0.6 \times 10^{5} cm^{-6} pc$. 
We obtained an estimate of $\rm 1.4 \times 10^{3} cm^{-3}$ for 
the electron density: $n_{e}$ $=
(EM/r)^{0.5}$, with $r$ being the core size which in this case corresponds
to the synthesized convolved beam size of $\rm
8\arcsec.7 \times 7\arcsec.8$. These values are on the lower side compared to
the estimates of Kurtz et al. (1994) obtained from their high-resolution maps at 8.3
GHz. They derive  values of $\rm 2\times 10^{6} cm^{-6} pc$ and 
$\rm 5.2 \times 10^{3} cm^{-3}$ for $EM$ and $n_{e}$, respectively. 
It should be noted here that the beam size of Kurtz et al. (1994) is much
smaller compared to the convolved beam size used in our derivation.

Taking the total integrated flux densities of 57 and 66 mJy estimated
from the 1280 and 610 MHz maps and using the formulation of Schraml \&
Mezger (1969) and the table from Panagia (1973; Table II), we estimate
the exciting zero age main-sequence star (ZAMS) 
of this UC HII region to be of spectral type
B0.5 -- B0. This is consistent with the estimates of Kurtz et al.
(1994; 1999). However, the FIR luminosity of $\rm 7.0\times10^{4}L_{\odot}$
from IRAS PSC implies an exciting star of spectral type O8 (Kurtz et
al. 1994). Radiative transfer modeling of the FIR and MIR
flux densities of IRAS 20178+4046 by Verma et al. (2003) also suggests an
O7 ZAMS exciting star.
In a later section (see Sect.\,\,\ref{mod.sect}), 
a detailed model of the emission
from the ISM around IRAS 20178+4046 has been presented, which implies 
the region to be powered by a cluster with the  
most massive star being of type B0. 
  
\subsection{Embedded cluster in the near-infrared}
\label{nir}

\begin{figure}
\resizebox{\hsize}{!}{\includegraphics{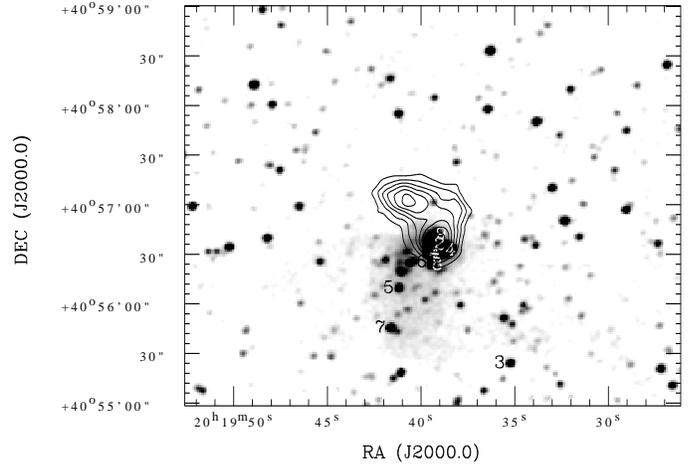}}
\caption{The 2MASS $K_{s}$-band image is shown. Overplotted are the
massive stars within the diffuse emission associated with the UC HII
region IRAS 20178+4046 (see discussion on the CM diagrams) 
The plus sign represents the IRAS point source
position. Contour map of
the dust emission at 850\,$\mu$m is also overlaid (see
Sect.\,\,\ref{jcmt.sect})}
\label{kband.fig}
\end{figure}

\begin{figure*}
\resizebox{\hsize}{!}{\includegraphics{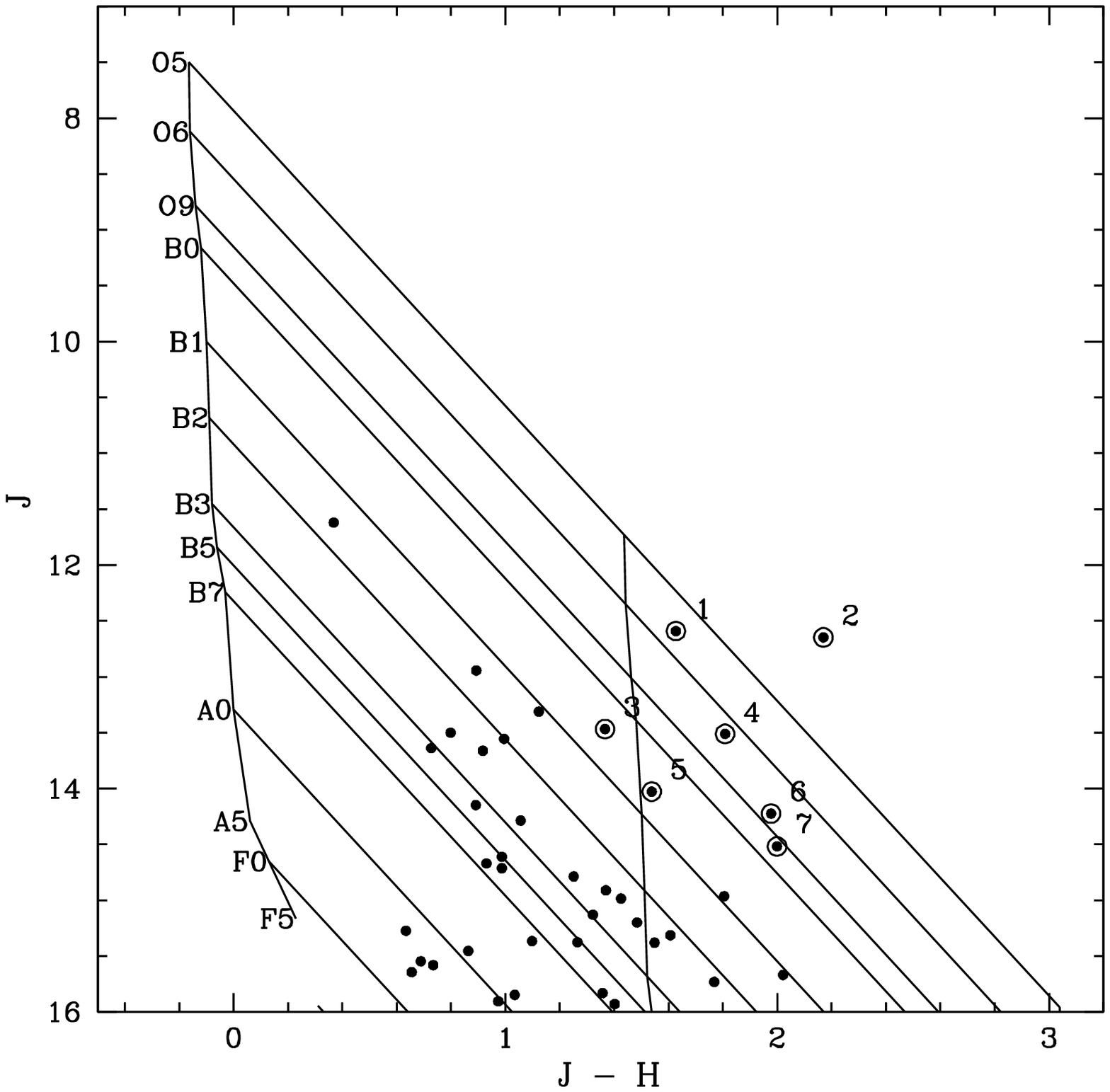}\includegraphics{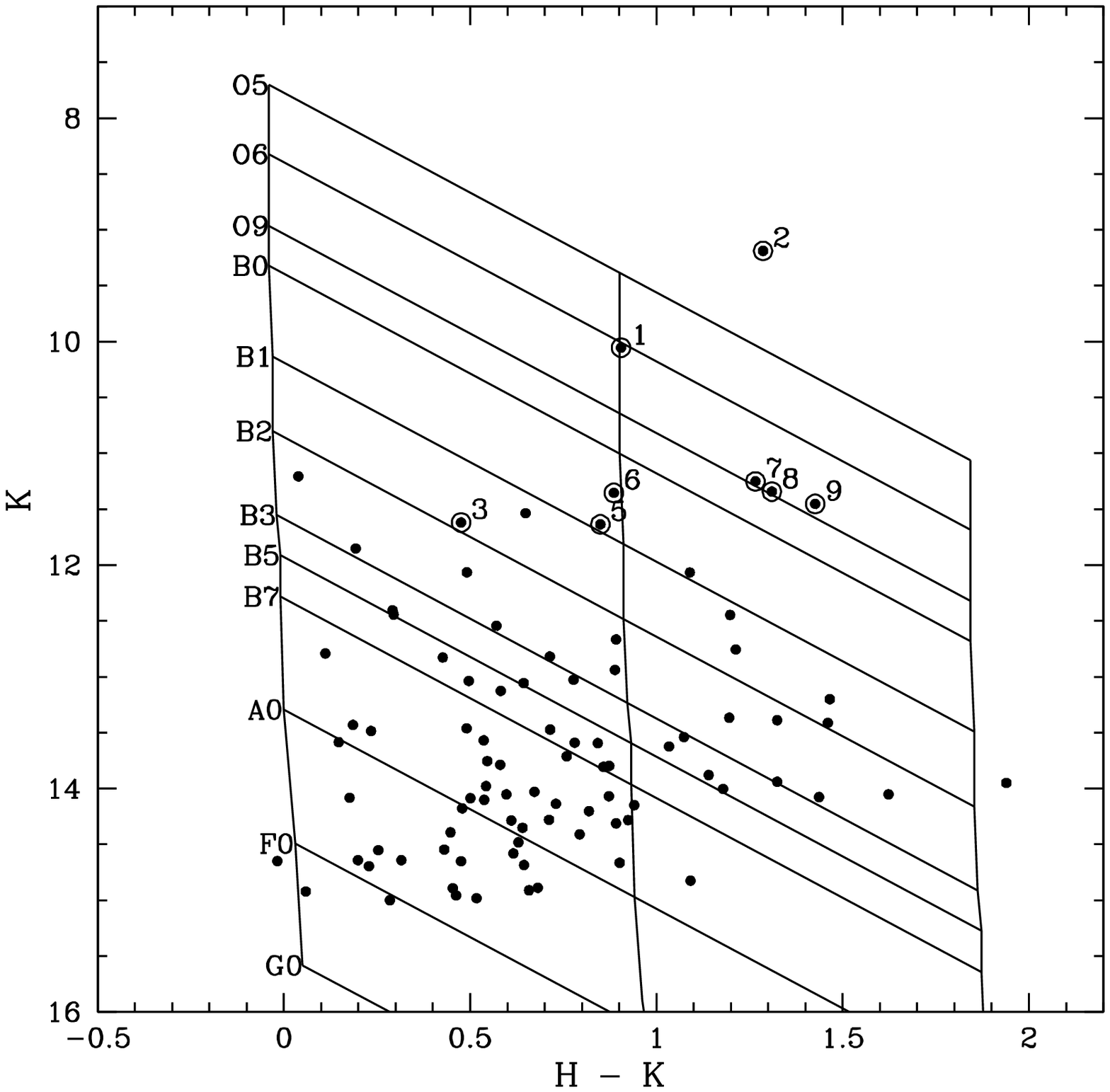}}
\caption{Colour-magnitude diagrams of the infrared cluster
in the IRAS 20178+4046 region. The nearly vertical solid lines
represent the zero age main sequence (ZAMS) loci with 0, 15, and 30 magnitudes of visual
extinction corrected for the distance. The slanting lines show the reddening vectors for each spectral
type. The magnitudes and the ZAMS loci are all plotted in the Bessell
\& Brett (1988) system.}
\label{cm.fig}
\end{figure*}

The 2MASS $K_{s}$ - band image of the region around IRAS 20178+4046
is shown in Fig.\,\,\ref{kband.fig}.
We see the presence of a stellar group/cluster mostly concentrated around
the UC HII region.
We use the 2MASS $JHK_{s}$ data to study the nature of these
sources seen in the vicinity of this UC HII region associated with IRAS
20178+4046. We select a large area of 90\arcsec\, radius centred on the
IRAS point source so as to completely cover the UC HII region and the
surrounding diffuse nebulosity seen in the near-infrared images. We
have restricted our sample of sources to those having good quality
photometry (2MASS `read-flag' 1 -- 3). 
Figure\,\,\ref{cm.fig} shows the two (($J-H$)/$J$ and ($H-K$)/$K$)
NIR colour-magnitude (CM) diagrams.
Using the ZAMS loci and the reddening vectors, we estimated 
the spectral type of the sources. 
The two CM diagrams show the presence of nine early type sources
with spectral types $\sim$ B0.5 or earlier, 
out of which six sources are common to both the diagrams.
Within an error of around one subclass (except for source \# 3), the
spectral type estimations derived from both the CM diagrams are
consistent.
It is likely that
the source \# 3 (which has the least extinction among the early
type stars) is a foreground star though it is not very close to
the zero extinction curve. It is also possible that it does not belong to
the group because spatially it is farther away (see Fig.\,\,\ref{kband.fig}). 
The nine early type sources identified from both the 
CM diagrams are listed in Table\,\,\ref{nir_src.tab}.
\begin{table*}
\caption{Early type sources ($\sim$ B0.5 and earlier)
 from the NIR CM diagram of IRAS 20178+4046}
\label{nir_src.tab}
\begin{tabular}{l|c|c|c|c|c}
\hline\hline
Source     & RA (2000.0) & DEC (2000.0) & J       & H       & K$_{s}$     \\
No.        & (hh:mm:ss.s)& (dd:mm:ss.s) & (mag)   & (mag)   & (mag)       \\
\hline
1         &20:19:39.5   &+40:56:30.5   & 12.491$\pm$0.034  &
10.929$\pm$0.023  &  10.016$\pm$0.027     \\
2         &20:19:39.3   &+40:56:35.9   & 12.532$\pm$0.030  &
10.433$\pm$0.039  &   9.150$\pm$0.029     \\
3         &20:19:35.2   &+40:55:24.0   & 13.380$\pm$0.022  &
12.076$\pm$0.019  &  11.580$\pm$0.015     \\
4         &20:19:38.7   &+40:56:32.6   & 13.400$\pm$0.040  &
11.659$\pm$0.034  &   --        \\
5         &20:19:41.2   &+40:56:09.6   & 13.930$\pm$0.026  &
12.456$\pm$0.018  &  11.597$\pm$0.024     \\
6         &20:19:40.5   &+40:56:25.3   & 14.118$\pm$0.046  &
12.209$\pm$0.032  &  11.315$\pm$0.029     \\
7         &20:19:41.6   &+40:55:45.5   & 14.405$\pm$0.037  &
12.475$\pm$0.030  &  11.212$\pm$0.026     \\
8         &20:19:39.4   &+40:56:23.7   &  --               &
12.610$\pm$0.029  &  11.304$\pm$0.025     \\
9         &20:19:39.2   &+40:56:42.4   &  --               &
12.833$\pm$0.032  &  11.414$\pm$0.019     \\
\hline
\end{tabular}\\ 
{\footnotesize Note: Magnitudes satisfying the `read-flag'
criteria of 1 -- 3 are given for the sources}\\
\end{table*}

\begin{figure}
\resizebox{\hsize}{!}{\includegraphics{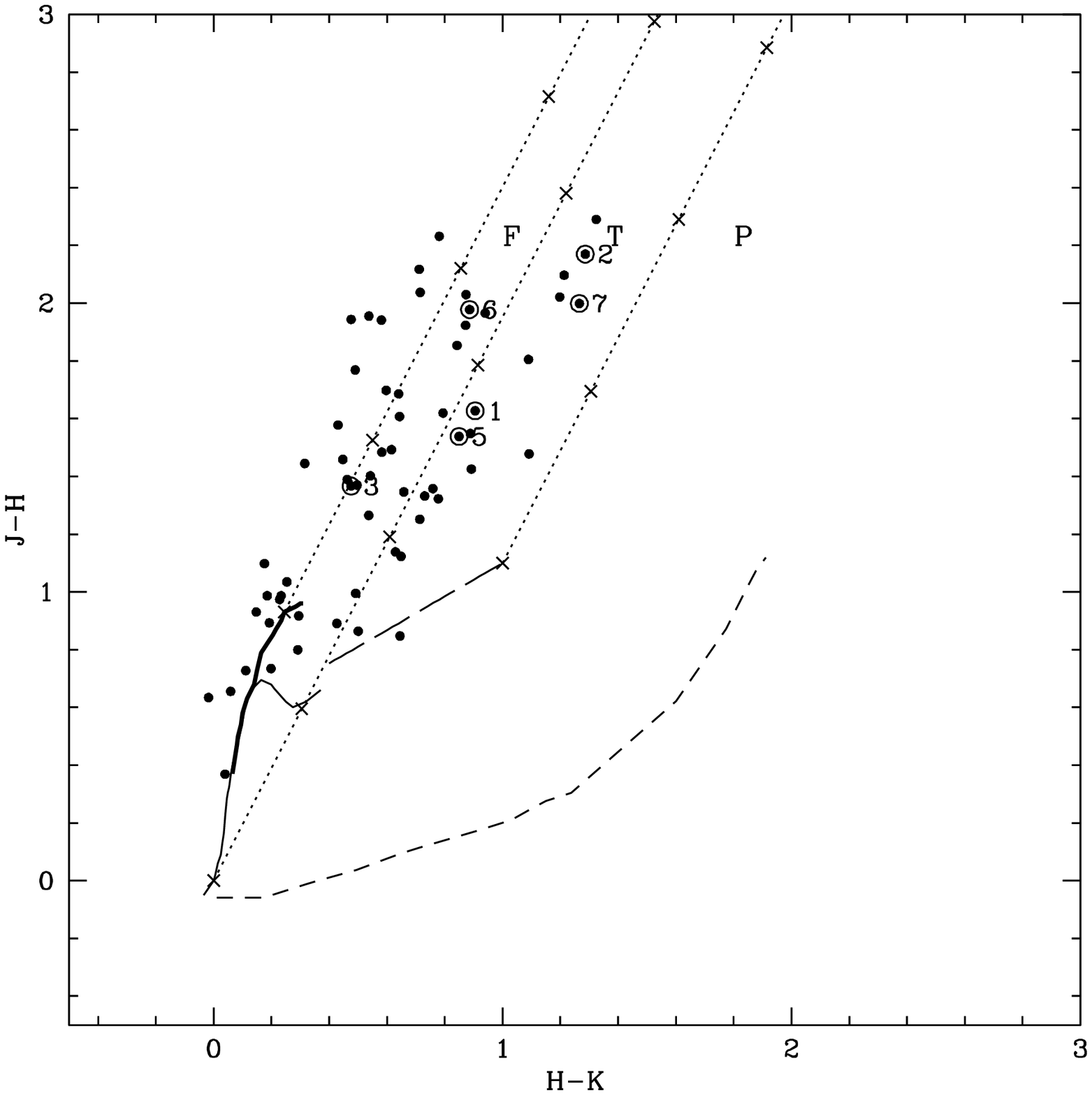}}
\caption{Colour-colour diagram of the sources in the vicinity of
the IRAS 20178+4046 region.
The two solid curves represent the loci of the main sequence (thin line) and the giant
stars (thicker line) derived from Bessell \& Brett (1988).
The long-dashed line is the classical T Tauri locus from
Meyer et al. (1997). The parallel dotted lines are reddening
vectors with the crosses placed along these lines at intervals of
five magnitudes of visual extinction. We have assumed the interstellar reddening law
of Rieke \& Lebofsky (1985) ($A_{J}/A_{V}$ = 0.282; $A_{H}/A_{V}$ =
0.175 and $A_{K}/A_{V}$ = 0.112). The short-dashed line represents the
locus of the Herbig AeBe stars (Lada \& Adams 1992). The plot
is classified into three regions namely ``F", ``T", and ``P" (see text
for details).
The colours and the curves shown in the figure are all transformed to the
Bessell \& Brett (1988) system.} 
\label{cc.fig}
\end{figure}

Figure\,\,\ref{cc.fig} shows the colour-colour (CC) diagram for the
sources in the region associated with IRAS 20178+4046. 
For clarity we have classified the CC diagram into three regions
(e.g. Tej et al. 2006; Ojha et al. 2004a \& b).
The sources in region ``F" are generally
considered to be either field stars, Class III objects, or Class II
objects with small NIR excess. The ``T" sources are classical
T Tauri stars (Class II objects) with large NIR excess or Herbig
AeBe stars with small NIR excess. The ``P" region has mostly 
protostar-like Class I objects and Herbig AeBe stars.  
Considering the quoted magnitude errors in 
Table\,\,\ref{nir_src.tab}, majority of 
the early type sources (except source \# 6 and \# 3) 
are seen to lie in the ``T" region.

In Fig.\,\,\ref{kband.fig}, we have overplotted sources 
listed in Table\,\,\ref{nir_src.tab} on the $K_{s}$-band 2MASS image.
It is interesting to see that five of these massive stars form a
cluster at the centre, of which the position of source \# 1 coincides
with the IRAS point source position. Comparing with the radio
morphology (Fig.\,\,\ref{1280_610.fig} left panel), sources
\# 1 and \# 2 are well within the compact ionized region. Sources \# 4,
\# 6, \#8, \# 9 envelope the UC HII region almost in a ring like
structure.
Apart from this clustering, three sources (\#6, \#5, \#7) form a
tail-type arc extending to the south. They lie within the diffuse
emission (southward lobe) seen in the $K_{s}$-band image.

The presence of the massive star clustering  makes the complex interesting 
because the radio flux density suggests a B0 -- B0.5 exciting source. 
It is likely that these sources
are deeply embedded. This results in absorption of a large fraction of
UV photons by the surrounding dust
which 
results in the underestimation of the spectral type from the radio flux density
of the ionized gas. Star \# 2  has spectral type much earlier
than O5 as indicated by both the CM diagrams. This could either be a
deeply embedded pre main-sequence (PMS) star or two or more unresolved
early-type stars. 
However, there is a discrepancy seen in the FIR
luminosity (Kurtz et al. 1994) when compared with the total luminosity
derived from the NIR sources. This could probably be
due to the distance uncertainties, IR excess and 2MASS resolution limit
which results in inaccuracies in the spectral type determinations
from the NIR CM diagrams. 
Hence, additional infrared spectroscopic observations are
essential in determining the exact nature and spectral type
of these sources.

\subsection{Dust emission in the sub-mm from JCMT-SCUBA data}
\label{jcmt.sect}
\begin{figure*}
\centering
\resizebox{\hsize}{!}{\includegraphics{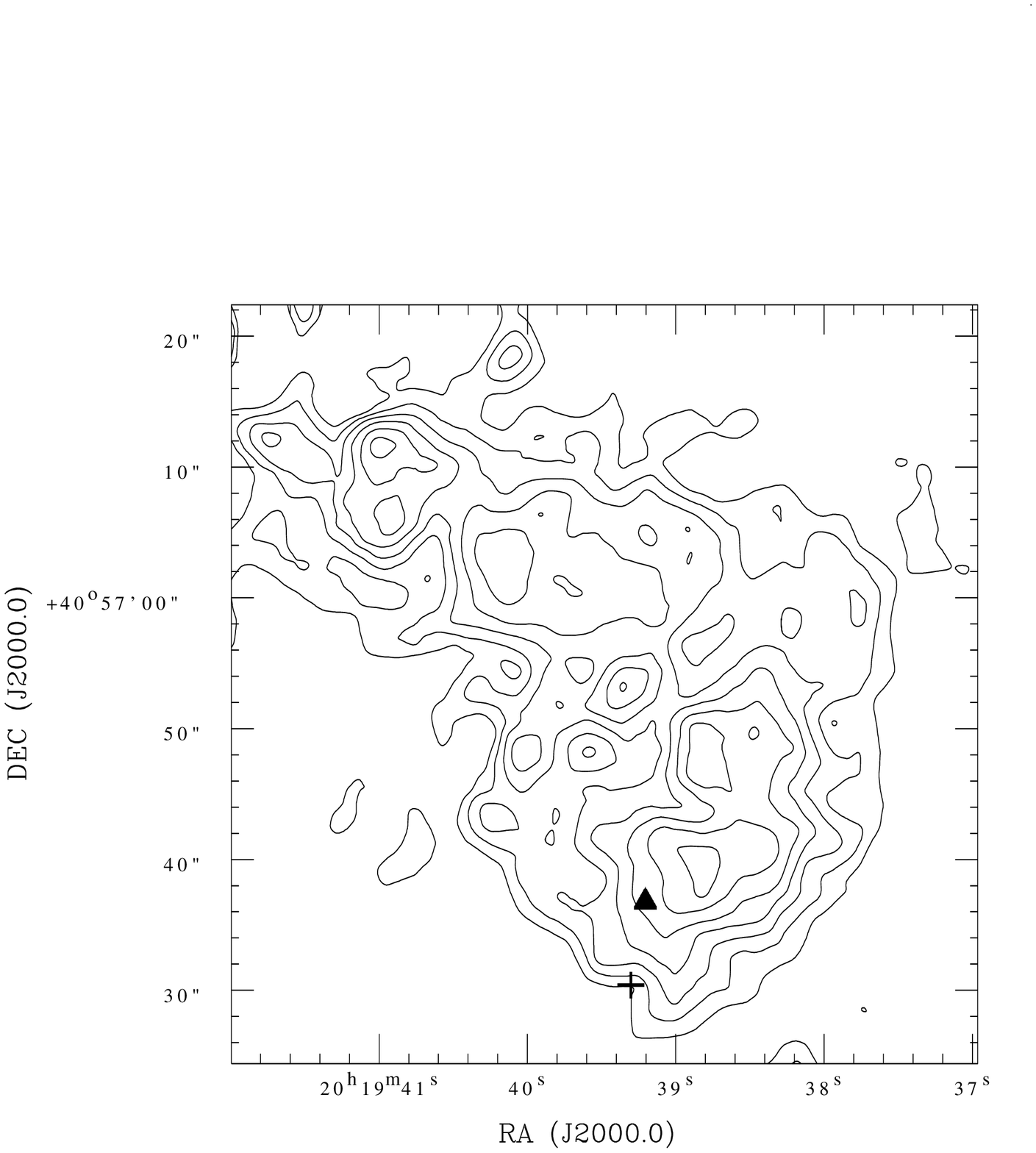} \includegraphics{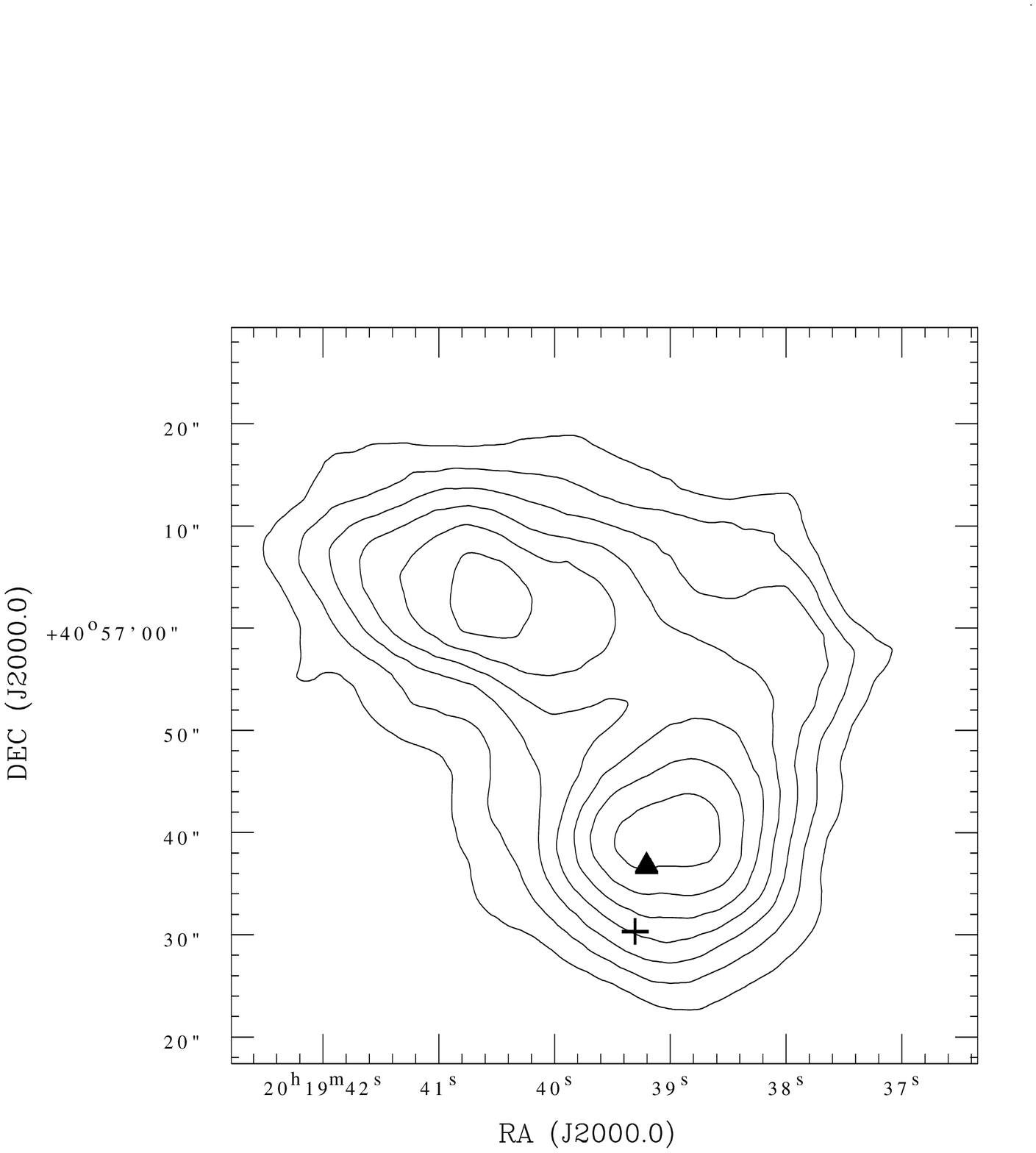}}
\caption{Contour maps showing the spatial distribution of dust emission at 450\,$\mu$m (left) and
850\,$\mu$m (right) for the region around IRAS 20178+4046.
The contour levels are at 35,45,55,65,75,85 and 95 \% of the peak
value of 4.19 and 1.72 Jy/beam at 450 and 850\,$\mu$m, respectively. 
The FWHMs of the symmetric 2-D Gaussian beams are
$\rm 8\arcsec.8$ and $\rm 14\arcsec.0$ for the two wave bands. The
plus sign and the filled triangle in each image mark the position of the IRAS point
source and the radio peak, respectively.}
\label{450_850.fig}
\end{figure*}

We used the emission in submillimetre wavebands to study the cold dust
environment in the region around IRAS 20178+4046. 
The spatial distribution of cold dust emission at 450 and 850\,$\mu$m
is displayed in Fig.\,\,\ref{450_850.fig}.
The angular resolutions are $\rm 8\arcsec.8$ and $\rm 14\arcsec.0$ for
the 450 and 850\,$\mu$m wave bands, respectively. The 450\,$\mu$m map
is relatively noisy and shows clumpy nature, whereas, the 850\,$\mu$m
map clearly shows the presence of two dust cores. 

The dust mass can be estimated from the following relation:
\begin{equation}
M_{dust} = 1.88 \times 10^{-4} \left(\frac{1200}{\nu}\right)^{3+\beta} S_{\nu}(e^{0.048\nu/T_{d}} - 1) d^{2}
\end{equation}
This is taken from Sandell (2000) and is a simplified version of Eq. 6 of
Hildebrand (1983). The above equation assumes the standard Hildebrand
opacities (i.e. $\rm \kappa_{1200GHz} = 0.1 cm^{2}g^{-1}$).
Here, $S_{\nu}$ is the flux density at frequency $\nu$,
$T_{d}$ the dust temperature, $\beta$
the dust emissivity index and is taken to be
2 (Hildebrand 1983), and $d$ the distance to the source in
kpc. We assume a typical dust temperature of 20 K (see Tej et al. 2006;
Klein et 2005). The flux densities are obtained from the JCMT-SCUBA maps shown in
Fig.\,\,\ref{450_850.fig}. To obtain the flux density of the entire cloud, we integrated up to the
last contour (which is at 35\% of the peak value).
Using the above relation, we estimate dust masses
of $\sim$ 7 and 15 $\rm M_{\odot}$
from the 450 and 850\,$\mu$m maps, respectively.
Assuming a gas-to-dust ratio of 100, the above values
translate to total masses of 700 and 1500 $\rm M_{\odot}$
for the cloud from the 450 and 850\,$\mu$m maps, respectively.
We estimate the individual masses of the these two dust cores to be $\sim$
250 (northern core) and 335 $\rm M_{\odot}$ (southern core) 
from the 850\,$\mu$m map. 
The cores are defined to cover regions upto $\sim$ 75\% of
the northern and southern intensity peaks.
Exploring the range of $T_{d}$ (20 -- 40
K) and $\beta$ (1 -- 2), we infer that the mass estimates can
be lower by up to a factor of $\sim$ 8.

\begin{figure}
\resizebox{\hsize}{!}{\includegraphics{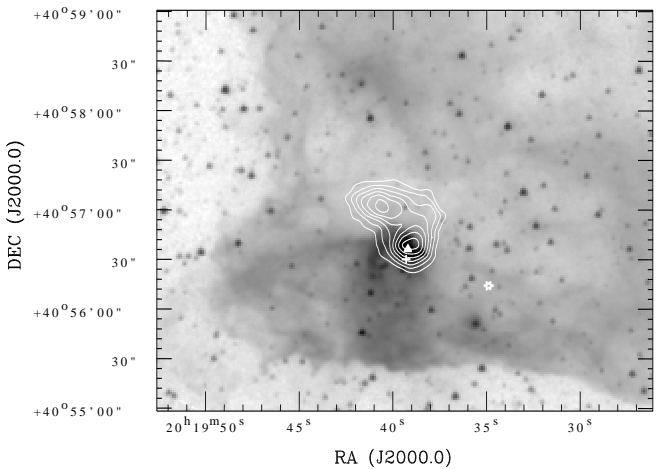}}
\caption{Contour map of dust emission at 850\,$\mu$m is
overlaid on the {\it Spitzer} 3.6\,$\mu$m band image for the region around IRAS
20178+4046. The filled triangle shows the position of the radio peak, the 
plus sign marks the position of the IRAS point source and the
star marks the position of the FIR peaks at 150 and
210\,$\mu$m (Verma et al. 2003).} 
\label{spitz_850_1280.fig}
\end{figure}

In Fig.\,\,\ref{spitz_850_1280.fig}, we present the various components of the
region associated with IRAS 20178+4046. The plot shows the contour map
of the dust emission at 850\,$\mu$m and the peak position of ionized gas
emission at 1280 MHz overlaid on the 3.6\,$\mu$m {\it Spitzer}-IRAC\footnote{The archival
image presented is the Post-Basic Calibrated Data (PBCD) downloaded using
the software Leopard. This work is based in part on observations made with
the {\it Spitzer Space Telescope}, which is operated by the Jet Propulsion
Laboratory, under NASA contract 1407.} image.
As compared to the $K_{s}$-band image, the diffuse emission seen in
the {\it Spitzer} image is far more extended.
The outer envelope spreads out to a ``hat-shaped" morphology.
It is interesting to note that the southern core of the 850\,$\mu$m
map coincides with the radio continuum emission and the cluster of
massive stars. 
One MIR source  (G078.4373+02.6584 -- $\rm \alpha_{2000.0} = 20^{h} 19^{m}
39^{s}.4\,;\,\delta_{2000.0} = +40^{\circ} 56\arcmin
35\arcsec$.2), lies inside the southern core coincident with the radio
peak position.
The MSX MIR colours $\rm F_{21/8}$, $\rm F_{14/12}$, $\rm F_{14/8}$,
and $\rm F_{21/14}$ of this source fall in the zone mostly occupied
by compact HII regions (Lumsden et al. 2002).
MIR emission in UIBs due to PAH (see Appendix A) 
is also seen from this region. The morphology of the UIB emission is
similar to that seen in the {\it Spitzer} image. 
Comparing with the FIR
intensity maps of Verma et al. (2003), it is seen that this
southern core is not spatially coincident with the FIR
peaks at 150 and 210\,$\mu$m but lies close to them 
(the core is located close to the last contours of the maps
presented in Fig. 5 of Verma et al. 2003). 
It is likely that this is an evolved
protocluster or a young (partly embedded) cluster where the 
massive stars develop UC HII regions and the cluster 
emerges out of the parental cloud and the
stars are detected in the near-infrared. Whereas,
the northern core is only detected in the sub-mm. No radio emission
(down to the level of the rms noise in the radio maps), MIR or NIR emission
is detected for this core. As is seen clearly from
Figs.\,\,\ref{kband.fig} and \ref{spitz_850_1280.fig}, the northern
core appears as an absorption region with a notable lack of stars.
This northern core is relatively further 
away from the emission peaks in the two FIR bands (150 \& 210\,$\mu$m) 
as compared to the southern
core. The mass of the two cores are similar and
satisfy the mass limit criteria of 100 $\rm M_{\odot}$ (Klein et al.
2005) for identifying earliest stages of massive star formation. 
The above scenario suggests that
this dense core could be harbouring a possible pre-protocluster or an
early protocluster candidate where we are sampling the initial phases
of cloud collapse in which massive stars have possibly begun to form
deeply embedded in the cluster.
Though the FIR data is not conclusive, the location of
the sub-mm dust cores with respect to the other components
most likely implies that we are seeing different
evolutionary stages of star formation in the two cores. 
A similar massive star forming
complex IRAS 06055+2039 was studied by Tej et al. (2006) where the dense
cloud core was at an earlier evolutionary stage compared to the associated HII region.

\subsection{Line emission from the Gas}
\label{mod.sect}

Spectroscopic signatures from the interstellar gas around 
IRAS 20178+4046 have been considered to explore additional
information about this star forming region. 
Millimeter-wave observations of CS($J$=2-1) by Bronfman et al (1996),
$^{13}$CO(2-1) by Wilkins et al (1989) and 
$^{13}$CO(1-0) by McCutcheon et al (1991)
has been translated to estimated gas mass of 544, 1130 and 1760 M$_{\odot}$
respectively, under reasonable assumptions. The lower value from 
the CS data is expected due to a smaller beam size
(39\arcsec) used.
These masses are similar to the estimates from the thermal
emission from cold dust, 700 -- 1500 M$_{\odot}$ 
(see Sect.\,\,\ref{jcmt.sect}). 
The available infrared spectroscopic measurements for IRAS 20178+4046
includes IRAS Low Resolution Spectrometer data
(LRS; covering 8 to 22 $\mu$m) and those 
by Faison et al. (1998) (covering 3 -- 5.3 \& 8 -- 13 $\mu$m).
The latter has detected the [Ne II] line at 12.8\,$\mu$m.
We have analysed the IRAS-LRS data to extract possible spectral 
signatures for a few selected ionic fine-structure lines   
commonly observed in HII regions. This resulted in clear detection of
the [Ne II] line and marginal detections of the lines 
due to [S III] (18.7 $\mu$m) and [Ar III] lines (8.99 $\mu$m).

\begin{figure}
\resizebox{\hsize}{!}{\includegraphics{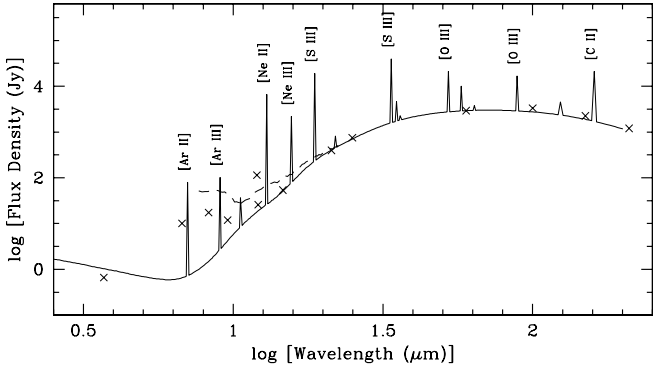}}
\caption{The emerging spectrum (solid line) predicted by the model for IRAS
20178+4046. The dashed line represent the IRAS LRS from Volk \& Cohen
(1989). The crosses represent the photometric data taken 
from Verma et al. (2003) except for the 4 MSX bands which were estimated from 
respective maps.}
\label{cloudy.fig}
\end{figure}

We have explored a self-consistent model of 
IRAS 20178+4046 region capable of explaining the observed infrared 
line emission as well as radio continuum from the photo-ionized gas.
The scheme developed by Mookerjea \& Ghosh (1999) has been used,
which is based on the photoionization code CLOUDY (Ferland 1996).
It considers various relevant physical and chemical 
processes involving constituents of the interstellar cloud.
The geometrical parameters describing the cloud were taken from 
Verma et al. (2003), who have successfully modelled the observed 
infrared SED and angular sizes using radiative transfer through
the dust component. 
Our best model corresponds to a constant density cloud of radial
dust optical depth, $\tau_{100{\mu}m}$ = 0.005, and a 
centrally embedded star cluster with upper mass cut-off of 
17.5 $M_{\odot}$ (B0 type star)
and the slope of the initial mass function of -1.6.  
The predicted radio emission at 1280 MHz is 51 mJy in very good agreement
with our GMRT measurement (57 mJy).
The emerging infrared spectrum including high-contrast lines
are presented in Fig.\,\,\ref{cloudy.fig} alongwith the
observed SED.
The inferred luminosity of the [Ne II] line 
from the measurements of Faison et al. (1998) and IRAS LRS
correspond to $\sim$ 74 \& 5.3 L${}_{\odot}$, respectively.
This discrepancy could be due to the very different beams and
spectral resolutions employed. 
The prediction of our model lies in between these two values, 
viz., 26.1 L${}_{\odot}$.
The luminosities for the [S III] and [Ar III] lines 
extracted from IRAS LRS are 7.7 \& 0.7 L${}_{\odot}$ respectively,
which are lower than those predicted.
However, the predicted line ratio [S III]/[Ne II] is very
similar to that from the LRS. 
Better measurements of these and other spectral lines are required 
to resolve the above issues. 
Several of the lines predicted 
(Fig.\,\,\ref{cloudy.fig}),
can easily be detected by the InfraRed
Spectrograph (IRS) instrument onboard  {\it Spitzer Space Telescope},
leading to better understanding of the star forming complex 
associated with IRAS 20178+4046. 
 
%
\section{General star formation scenario in IRAS 20178+4046}
\label{starform.sect}

The morphology of the radio emission of this UC HII
region at higher resolution (Kurtz et al. 1994) is cometary
in shape.
As mentioned in Garay \& Lizano (1999), the champagne flow
and the bow shock models explain the cometary morphologies
satisfactorily though observationally it is difficult to
differentiate between the two. The morphology of the various
components presented in Fig.\,\,\ref{spitz_850_1280.fig}
seems to suggest a champagne flow. In this model the medium
in which the massive star is born has strong density
gradients which results in the HII region moving
supersonically away from the high density region in a
so-called champagne flow (Garay \& Lizano 1999). The HII
region is density bounded and the pressurized HII gas bursts
out of the cloud into a fan-like or plume-like structure. The presence
of the dense dust core and the high density to the
north of the UC HII region and the fan-like morphology seen
in the {\it Spitzer} image and the MIR image 
adds strength to this conjecture. It is important to note that
the fan-like morphology is seen in the other three IRAC bands
also (4.5, 5.8 and 8\,$\mu$m which are not presented in this
paper). Theoretical
simulations in the champagne phase by Yorke et al. (1983)
show that during this phase the resulting configuration is
ionization bounded on the high-density side and density
bounded on the side of the outward flow. The {\it Spitzer} image
shows the presence of another high density lane to the south
of the fan-like diffuse emission. This southern high density
lane is spatially coincident with the east-west extended
emission detected in the low resolution 8.3 GHz map of Kurtz
et al. (1999). Further high resolution observations of molecular 
gas velocity structure in this star forming region is crucial 
in understanding the details of the environment and formation
mechanism of the massive stars. 
Inspite of the intrinsic temporal
variability typically seen in masers (Stahler \&
Palla 2004) and the sensitivity of the
single dish observations carried out for this source, 
the absence of the OH maser could possibly be an
indication of a more evolved HII
region (Codella et al. 1994) in support of the optically thin scenario
infered from our radio observations.
\section{Conclusions}
\label{concl.sect}
The massive star forming region associated with IRAS 20178+4046 
has been studied in detail in the radio, infrared and
sub-mm wavelengths, leading to the following conclusions 
\begin{itemize}
\item[1.]
High-sensitivity and high-resolution radio continuum maps at 1280 and
610 MHz obtained from our observations using GMRT show a simple compact
spherical morphology. The total integrated emission implies an exciting star of spectral
type B0 -- B0.5.
\item[2.]
The sub-mm emission from cold dust has been studied
using JCMT-SCUBA at 450 and 850\,$\mu$m, from which 
we estimate the total mass of the cloud to be 
$\sim$ 700 -- 1500 $\rm M_{\odot}$ which is
also supported by the estimates derived from the
millimeter-wave molecular line data.
The 850\,$\mu$m map shows the presence of two dense cores
of masses $\sim$ 250 and 335 $\rm M_{\odot}$.
The southern core is most likely an evolved protocluster or a young cluster
where the massive stars have formed the UC HII region and the cluster emerges out
of the parental cloud. The northern core is only detected in the sub-mm and is a
possible pre-protocluster candidate.
\item[3.]
The NIR colours from 2MASS data suggest the presence of 
several massive stars (earlier than $\sim$ B0.5) within and enveloping the 
UC HII region which are likely to be deeply embedded.
\item[4.] 
The embedded star cluster has been
characterized by modelling the observed radio
continuum and infrared SED using a photoionization
code. The upper mass cut-off is found to be 17.5$\rm
M_{\odot}$ (B0 type star) for an usual initial mass
function.
\end{itemize}

\subsection*{Acknowledgments}
We thank the anonymous referee for providing critical comments and
suggestions, which have helped in improving the scientific content
of this paper. We also thank Dr. Malcolm Walmsley for his
useful suggestions.
We thank the staff at the GMRT who have made the radio observations
possible. GMRT is run by the National Centre for Radio Astrophysics of
the Tata Institute of Fundamental Research.

\begin{appendix}
\section{Spatial distribution of UIBs from the MSX data}
\label{msx.sect}
\begin{figure}
\resizebox{\hsize}{!}{\includegraphics{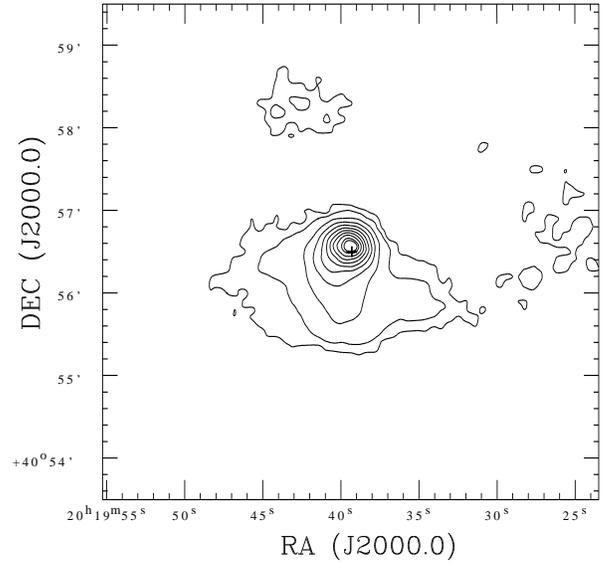}}
\caption{Spatial distribution of the total radiance in the UIBs 
for the region around IRAS
20178+4046 as extracted from the MSX four band images. The contour
levels are at 5, 10, 20, 30, 40, 50, 60, 70, 80, 90, and 95 \% of
the peak emission of $\rm 1.5\times10^{-4}$ W\,m$^{-2}$\,Sr$^{-1}$.}
\label{msx.fig}
\end{figure}

We have used the scheme developed by Ghosh \& Ojha (2002) 
to extract the contribution of UIBs (due to the PAHs) 
from the mid-infrared images in the four MSX bands. 
The scheme models the observations with a combination of thermal
emission from the `normal' large interstellar dust grains (gray body)
and the UIB emission from the gas component, under reasonable 
assumptions.
The spatial distribution of emission in the UIBs
with an angular resolution of $\sim$ 18$^{\prime\prime}$ 
(for the MSX survey) extracted
for the region around IRAS 20178+4046 is shown in 
Fig.\,\,\ref{msx.fig}. 
The morphology of the UIB emission is similar to that of the diffuse
emission seen in the $K_{s}$-band and {\it Spitzer} images shown in
Figs.\,\,\ref{kband.fig} and \ref{spitz_850_1280.fig}. 
The integrated
emission (upto 5\% of the peak value) from the UIB features 
in band A (viz., 6.2, 7.7, 8.7\,$\mu$m)
and band C (11.3, 12.7\,$\mu$m) is found to be $\rm 7.54\times10^{-12}
Wm^{-2}$. 
In comparison, the emission in
the PAH bands (6.2, 7.7 \& 11.3\,$\mu$m) from the ISOCAM 
measurements is reported to be $\rm 1.87\times10^{-12} Wm^{-2}$ 
in Verma et al. (2003). 
Our larger value is not surprising considering that it includes
two additional features and also covers a larger area.
In addition, spatial distribution of warm dust component
(optical depth; $\tau_{10{\mu}m}$) and its temperature ($\rm T_{MIR}$) have
been generated (maps not presented here).
The distribution of $\tau_{10{\mu}m}$ is compact with a peak
value of 0.018, the peak coinciding with that of the UIB
emission. On the other hand, $\rm T_{MIR}$ distribution is more extended
ranging between $\sim$ 80 -- 193 K, with the warmest region
to the western edge of extended UIB emission.  

\end{appendix}

\end{document}